\documentclass[reprint, superscriptaddress, secnumarabic, amssymb, nobibnotes, aps, prl]{revtex4-1}

\usepackage{chemformula,siunitx}
\usepackage{lastpage}
\usepackage[protrusion=true, expansion=true]{microtype}
\usepackage{wrapfig}
\usepackage{multirow}
\usepackage{array}
\usepackage{ragged2e}
\justifying

\usepackage{graphicx}
\usepackage{epstopdf}
\usepackage[T1]{fontenc}
\usepackage{amsbsy}
\usepackage[T1]{fontenc}
\usepackage{amsmath}
\usepackage{amssymb}
\usepackage{bbm}
\usepackage{braket}
\usepackage{xcolor}
\allowdisplaybreaks
\usepackage{graphicx}

\usepackage[normalem]{ulem}

\newcommand{\equref}[1]{Eq.~(\ref{#1})}

\newcommand{\figref}[1]{Fig.~\ref{#1}}

\newcommand{\tableref}[1]{Table~\ref{#1}}

\usepackage[colorlinks=true]{hyperref}
\hypersetup{
    unicode=false,          
    pdftoolbar=true,        
    pdfmenubar=true,        
    pdffitwindow=false,     
    pdfstartview={FitH},    
    pdftitle={ZrTe2.8},    
    pdfauthor={Poulami Manna},      
    pdfsubject={},   
    pdfcreator={},   
    pdfproducer={}, 
    pdfkeywords={} {} {}, 
    pdfnewwindow=true,      
    colorlinks=true,       
    linkcolor=blue, 
    citecolor=blue,        
    filecolor=magenta,      
    urlcolor=blue          
}

\begin{document}

\title{\textrm{Evidence of anisotropic bulk superconductivity in disorder-induced ZrTe$_{3-x}$}}

\author{P. Manna}
\affiliation{Department of Physics, Indian Institute of Science Education and Research Bhopal, Bhopal, 462066, India}
\author{C. Patra}
\affiliation{Department of Physics, Indian Institute of Science Education and Research Bhopal, Bhopal, 462066, India}
\author{T. Agarwal}
\affiliation{Department of Physics, Indian Institute of Science Education and Research Bhopal, Bhopal, 462066, India}
\author{S. Srivastava}
\affiliation{Department of Physics, Indian Institute of Science Education and Research Bhopal, Bhopal, 462066, India}
\author{S. Sharma}
\affiliation{Department of Physics, Indian Institute of Science Education and Research Bhopal, Bhopal, 462066, India}
\author{P. Mishra}
\affiliation{Department of Physics, Indian Institute of Science Education and Research Bhopal, Bhopal, 462066, India}
\author{R.~P.~Singh}
\email[]{rpsingh@iiserb.ac.in}
\affiliation{Department of Physics, Indian Institute of Science Education and Research Bhopal, Bhopal, 462066, India}

\begin{abstract}
Transition-metal trichalcogenides distinguish themselves from other two-dimensional materials in nanoscience and materials science due to their remarkable range of intrinsic properties, including various electronic, optical, and magnetic behaviors. Here, we report a comprehensive study of superconductivity in disordered ZrTe$_{3-x}$ ($x$ = 0.2) with suppressed charge density wave. We observe a type-II bulk anisotropic superconductivity with a superconducting transition at $T_c$ = 3.59(4) \si{K}. Angle-dependent upper critical field measurements and Berezinskii-Kosterlitz-Thouless transition confirm the inherent quasi-two-dimensional nature of superconductivity in this disordered system.
\end{abstract}

\maketitle

\section{INTRODUCTION}
The complex interplay between charge density waves (CDWs) and superconductivity, characterized by competing \cite{doi:10.1126/science.1223532, PhysRevB.89.224513} and coexisting \cite{doi:10.1143/JPSJ.79.123710}, has posed intriguing questions in condensed matter physics. These correlated phenomena are mainly caused by Fermi-surface instabilities and electron-phonon interactions \cite{RevModPhys.60.1129}. In the weak-coupling regime, they often compete for the same electronic states, so that introducing disorder can suppress one order while simultaneously enhancing the other \cite{petrovic2016disorder, PhysRevLett.122.017601}. While this interplay has gained significant traction through studies of cuprate superconductors \cite{doi:10.1126/science.1223532, chang2012direct, le2014inelastic}, it is equally prevalent in transition-metal trichalcogenides and layered transition-metal dichalcogenides \cite{zhu2016superconductivity, lian2023interplay}.

Group IVB transition-metal trichalcogenides (TMTCs), distinguished by their strongly anisotropic crystal structures, form a notable subclass of two-dimensional (2D) materials \cite{patra2020anisotropic}. Their unique combination of structural and electronic features makes them ideal candidates for exploring emergent physical phenomena, particularly in addressing the long-standing puzzle surrounding the interplay between dimensionality and superconductivity. They additionally offer remarkable quasi-one-dimensional (quasi-1D) characteristics, which combine the advantages of 2D materials with promising platforms for fabricating nanoelectronic devices and probing new avenues in nanotechnology. Different compositions of binary zirconium tellurium compounds have garnered significant research interest due to their intriguing properties, including negative magnetoresistance \cite{doi:10.1021/acsnano.9b02196, wang2022magnetotransport}, quantum Hall effect \cite{tang2019three}, log-periodic oscillations \cite{doi:10.1126/sciadv.aau5096}, and potential topological characteristics \cite{PhysRevB.94.165201, bouhon2020non}. 

In this context, the CDW compound ZrTe$_3$ \cite{PhysRevB.80.075423} is of particular interest due to its unusual coexistence of filamentary and bulk superconductivity \cite{PhysRevB.85.184513}. ZrTe$_3$ crystallizes in the TaSe$_3$-type structure with the space group P2$_1$/m (11). Along the $b$ direction, it consists of an infinite quasi-1D chain, whereas quasi-2D Zr-Te layers are formed along the $ac$-plane \cite{PhysRevB.91.155124}. Pristine ZrTe$_3$ undergoes a CDW transition around 63 K, driven by Fermi surface nesting \cite{PhysRevB.80.075423} with filamentary superconductivity emerging only below $2 \text{ K}$ \cite{gu2018pressure}. While bulk superconductivity can be induced via elemental substitution, atomic intercalation, disorder tuning, or the application of external pressure \cite{DEFARIA2024175919, ISHIKURA2026131145, yamaya2002effect, PhysRevB.71.132508, wang2023superconductivity, PhysRevLett.106.246404, Lei_2011}, these methods often obscure intrinsic electronic properties, and the exact nature of superconductivity remains elusive. Recent research demonstrates that suppressing the CDW state via controlled disorder can induce robust bulk superconductivity with an enhanced transition temperature ($T_c$), providing a clearer avenue for probing the fundamental physics of quasi-2D superconducting systems \cite{patra2024superconducting, patra2024planar, PhysRevB.87.024508, manna2025quasi, agarwal2023quasi}.


Here, we report the successful synthesis of crystalline ZrTe$_{3-x}$ ($x$ = 0.2) and comprehensive characterization of its superconducting properties via magnetization, specific heat, and transport measurements. The results confirm anisotropic type-II superconductivity with $T_c$ = 3.59(4) \si{K} and complete suppression of the CDW state. Furthermore, angle-dependent magnetotransport and clear signatures of a Berezinskii-Kosterlitz-Thouless (BKT) transition provide compelling evidence for quasi-two-dimensional superconductivity in bulk single crystals of ZrTe$_{3-x}$ ($x$ = 0.2).

\begin{figure*}
\includegraphics[width=2.05\columnwidth]{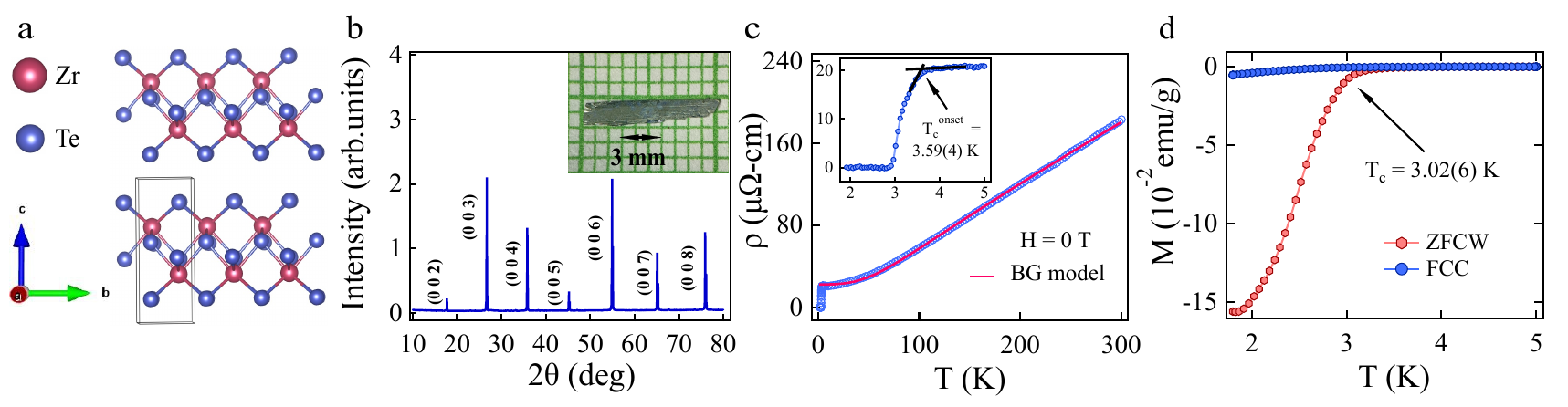}
\caption {\label{fig1}(a) The crystal structure of ZrTe$_{2.8}$ is shown, where the pink and blue spheres stand in for the Zr and Te atoms, respectively. (b) Distinct $(00l)$ reflections are visible in single-crystal XRD patterns, suggesting a preferential orientation along the c-axis. Inset: Crystal image taken with an optical microscope. (c) A magnified view of the resistivity near the superconducting transition reveals an onset transition temperature of approximately 3.59(4) K. Inset: The temperature-dependent electrical resistivity conducted in zero magnetic field is fitted with the BG model. (d) Magnetization versus temperature measurements under a very low applied field display a broad superconducting transition with a critical temperature of 3.02(6) \si{K}.}
\end{figure*}

\section{EXPERIMENTAL DETAILS}
ZrTe$_{2.8}$ single-crystals were grown employing the typical isothermal chemical vapor transport (ICVT) method, with iodine (I$_2$) chunks serving as the transport agent. Zr (5N) lumps and finely powdered Te (4N) pieces were combined in accurate stoichiometric proportions and sealed in an evacuated quartz tube along with \ch{I_2} (5mg/cc). The tube was then placed in a box furnace and kept at a constant temperature of 900 \si{K}. After a 10-day reaction time, the system was cooled in ice-water and large shiny crystals formed. The identity and phase purity of the crystals were examined using powder X-ray diffraction (PXRD) applying monochromatic Cu-$K_\alpha$ radiation ($\lambda$ = 1.54 \AA) on an X'pert PANalytical Empyrean X-ray diffractometer, and the Laue diffraction pattern was obtained using a Photonic Science Laue camera. Elemental compositions were confirmed by energy-dispersive X-ray analysis (EDAX) in combination with scanning electron microscopy (SEM). Magnetization measurements were performed on a Quantum Design magnetic property measurement system (MPMS3) equipped with a cryostat $^4He$. A 9T-Quantum Design physical property measuring system (PPMS) was introduced to carry out transport and specific heat measurements using conventional four-probe and two-tau methods, respectively.

\section{RESULTS AND DISCUSSION}
\subsection{Sample characterization}
\figref{fig1}(a) displays the crystal structure of ZrTe$_{2.8}$ along the $a$-axis, produced by VESTA software \cite{momma2011vesta}. The room-temperature powder X-ray diffraction patterns of the crushed crystal of ZrTe$_{2.8}$ were analyzed using the Rietveld refinement \cite{fullprof}, shown in (a), which confirms a single phase present in Te-deficient ZrTe$_{3-x}$. This compound has a monoclinic structure with space group P2$_1$/m (11). The lattice parameters obtained from the refinement are $a$ = 5.9052(7) $\text{\AA}$ $b$ = 3.9492(6) $\text{\AA}$, $c$ = 10.1121(5) $\text{\AA}$, which are slightly higher than the parent compound. The inset of \figref{fig1}(b) shows the microscopic image of a 1 cm long silver-colored crystal. The single crystal XRD patterns of ZrTe$_{2.8}$ are illustrated in \figref{fig1}(b), where the crystals are oriented along the $(00l)$ direction, implying the $c$-axis as a growth axis.  The Laue diffraction pattern (Left inset of \textcolor{blue}S1(b), Supplemental Material \cite{Supp}) confirms the crystallinity of the sample, while the EDAX analysis verifies the intended elemental compositions (see Fig. \textcolor{blue}S1(b), Supplemental Material \cite{Supp}).
 
\begin{figure*}
\includegraphics[width=1.5\columnwidth]{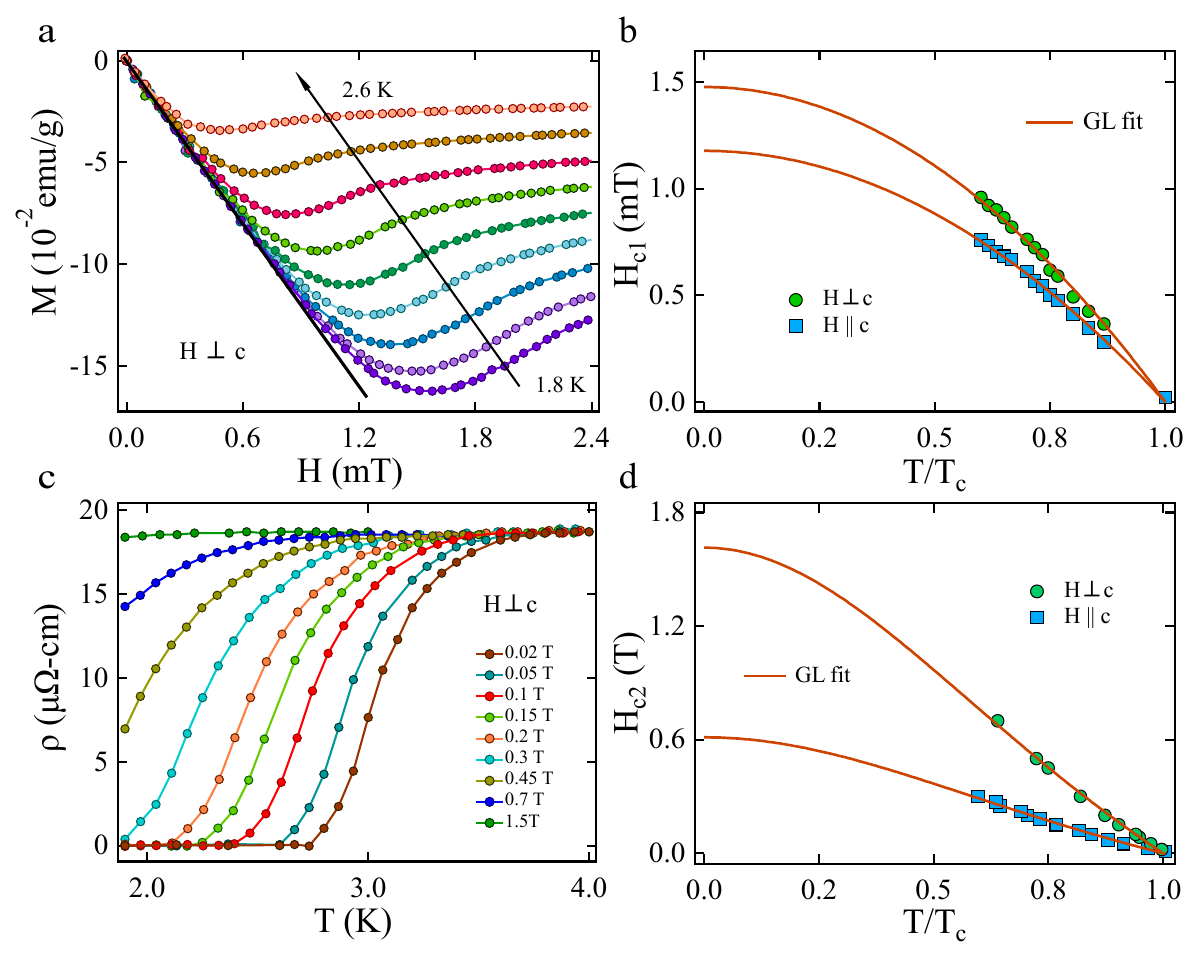}
\caption {\label{fig2}(a) Field-dependent magnetization curves at various temperatures for $H$ perpendicular to c-axis is shown. (b) Temperature-dependent lower critical fields in both directions. GL-fitting is shown by the orange lines. (c) Temperature-dependent electrical resistivity for magnetic fields applied parallel to the ab plane. (d) The conventional GL model is used to fit upper critical fields in both orientations as a function of reduced temperature, represented by orange solid lines.}
\end{figure*}

\subsection{Electrical resistivity and Magnetization}
The temperature-dependent AC electrical resistivity of the crystalline ZrTe$_{2.8}$ was measured from 1.9 K to 300 K in the absence of an external magnetic field, as illustrated in \figref{fig1}(c). The gradual decrease in resistivity from 300 K to 5 K reflects the metallic behavior of the compound, with a residual resistivity ratio (RRR), $\rho(300 K)$/$\rho(5 K)$, of roughly 9, indicative of relatively low scattering from defects or impurities in the crystal. A broad superconducting transition is observed at low temperatures: the onset of superconductivity begins at a higher temperature of 3.59(4) K, while the resistivity drops to zero at 2.86(2) K, as illustrated schematically in the inset of \figref{fig1}(c). Interestingly, no resistivity anomaly is observed near 63 K, where a charge density wave (CDW) transition would normally be expected as the parent compound. It suggests suppression of the CDW phase, likely due to Te deficiency, which can weaken the nesting condition necessary for CDW formation. The normal-state resistivity is well described by the Bloch-Grüneisen (BG) model, which captures the temperature dependence of electron-phonon scattering. This model is expressed as 
\begin{equation}
{\rho(T)}= {\rho_{0}}+{\rho_{BG}(T)}.
\label{Eq1:Parallel}
\end{equation}
Where, ${\rho_{BG}(T)}$, the temperature-dependent contribution of BG is defined as
\begin{equation}
{\rho_{BG}(T)}= r \left(\frac{T}{\theta_{D}}\right)^3 \int_{0}^{{\theta_{D}}/T} \frac{x^{3}}{(e^{x}-1)(1-e^{-x})} dx.
\label{Eq2:Parallel}
\end{equation}
$\rho_{0}$ represents the temperature-independent residual resistivity, $r$ is a material-dependent constant, and $\theta_{D}$ denotes the Debye temperature. The fitting parameters are estimated as $\rho_{0}$ = 22.5(4) \si{\mu\ohm}-cm and the Debye temperature $\theta_{D}$ = 366(1) K.

Temperature-dependent magnetization measurements offer crucial evidence for identifying bulk superconductivity. These measurements were carried out using two standard protocols: zero-field-cooled warming (ZFCW) and field-cooled cooling (FCC), under an applied magnetic field of 1 \si{mT}. The superconducting transition temperature is typically determined from the bifurcation point of the ZFCW and FCC curves, as illustrated in \figref{fig1}(d). A distinct diamagnetic response observed in both modes confirms the onset of bulk superconductivity at 3.02(6) K. The comparatively weaker magnetization signal in the FCC mode relative to the ZFCW mode is likely due to strong flux pinning effects within the sample.

\begin{table}[b]
\caption{The anisotropic superconducting parameters.}
\label{tbl1}
\setlength{\tabcolsep}{13pt}
\begin{center}
\renewcommand{\arraystretch}{1.2}
\begin{tabular}[b]{lccc}\hline
Parameters& Unit & $H\parallel c$ & $H\perp c$ \\
\hline
$H_{c1}(0)$ & mT & 1.17(6) &   1.47(5) \\ 
$H_{c2}^{res}$(0) & T & 0.61(2) &   1.61(4) \\
$\xi$ & nm & 8.82(1) &  23.2(4)  \\
$\lambda_{GL}$ & nm & 620.2(2)  & 689.0(4)  \\
$\kappa_{GL}$& & 29.7 & 45.7  \\
\hline
\end{tabular}
\end{center}
\end{table}

To determine the lower critical field $H_{c1}$(0), field-dependent magnetization measurements were performed on a range of fixed temperatures. The $H_{c1}$ values were identified from the magnetization curves (M) versus the magnetic field (H) at the point where the data begin to deviate from the initial linear region, corresponding to the Meissner effect. This deviation marks the onset of magnetic flux penetration into the sample, signaling the lower critical field. As shown in \figref{fig2}(a) and \textcolor{blue}S2(a) (Supplemental Material \cite{Supp}), these measurements were performed for two configurations: with the magnetic field perpendicular and parallel to the crystallographic $c$-axis, respectively. The extracted $H_{c1}$(T) values were then analyzed as a function of reduced temperature employing the typical Ginzburg-Landau (GL) relation described as
\begin{equation}
H_{c1}(T)=H_{c1}(0)\left[{1-t^{2}}\right], \quad  \text{where} \;  t = \frac{T}{T_{c}}
\label{eqn3:HC1}
\end{equation}
which gives $H_{c1}$(0) = 1.17(6) and 1.47(5) mT for $H$ parallel to the $c$-axis and $H$ perpendicular to the $c$-axis, respectively, as given in \figref{fig2}(b). Temperature-dependent resistivity measurements were conducted under various externally applied magnetic fields to evaluate the upper critical field ($H_{c2}$(0)), as presented in \figref{fig2}(c) and \textcolor{blue}S2(b) (Supplemental Material \cite{Supp}) for magnetic fields aligned parallel and perpendicular to the same plane, respectively. The increase in the magnetic field leads to a systematic decrease in the superconducting transition temperature ($T_{c}$). We adopted a commonly used criterion based on the decrease in resistivity, where the transition point is defined as $\rho$ = 0.9$\rho_n$, with $\rho_n$ denoting the resistivity of the normal state just above the transition. The resulting extrapolated $H_{c2}$(T) data were then examined using the standard GL framework (\equref{eqn4:HC2}) that enables the extrapolation of $H_{c2}$(0).
\begin{equation}
H_{c2}(T) = H_{c2}(0)\left[\frac{1-t^{2}}{1+t^{2}}\right],  \quad  \text{where} \;  t = \frac{T}{T_{c}}.
\label{eqn4:HC2}
\end{equation}

So, the estimated values of $H_{c2}$(0) are 0.61(2) and 1.61(4) T for $H \parallel c$ and $H \perp c$, respectively (\figref{fig2}(d)). These fits not only provide the zero-temperature limit of $H_{c2}$ for both orientations, but also provide insight into the anisotropic nature of superconductivity, which can be quantified by calculating the anisotropic parameter $\Gamma$ = ${H_{c2}^{\parallel c}}$/${H_{c2}^{\perp c}}$ as 0.38. The moderate anisotropy observed in this system is consistent with the quasi-two-dimensional character of the superconducting state, as expected from layered structural motifs. A comparative overview of anisotropy parameters across bulk and monolayer superconductors is summarized in S2(d). This quasi-2D behavior is further supported by the angle-dependent upper critical field measurements, which reinforce the presence of anisotropic superconducting properties.

The Pauli limiting mechanism and the orbital limiting effect are two fundamental factors that predominantly govern the destruction of superconductivity in the presence of an external magnetic field. The Pauli limiting effect involves Zeeman splitting, which breaks Cooper pairs by aligning electron spins in the same direction. In contrast, the orbital limiting effect, originating from the Lorentz force, disrupts the coherent motion of Cooper pairs through the induction of vortices. Following the BCS theory, the Pauli limit effect is calculated as 6.67(7) T using $H_{c2}^{P}$(0) = 1.86 $T_{c}$ with $T_{c}$ = 3.59(4) K \cite{Chandrasekhar1962pauli, Clogston1962pauli}. This value lies above the experimentally determined in-plane upper critical field, inferring that the Pauli limit is preserved. Additionally, the orbital limiting effect is assessed using the Werthamer-Helfand-Hohenberg (WHH) theory, applicable to type-II superconductors under the assumption of negligible spin-orbit coupling \cite{WHH1966orbital, Helfand1966orbital}. It is defined as 
\begin{equation}
H^{orb}_{c2}(0) = -\alpha T_{c} \left.{\frac{dH_{c2}(T)}{dT}}\right|_{T=T_{c}}. 
\label{eqn5:WHH}
\end{equation}
Here, $\alpha$ is a dimensionless constant referred to as the purity factor, which distinguishes between superconductors of the clean and dirty limit, taking values of 0.73 and 0.69 for superconductors of the clean and dirty limit, respectively. The estimated value of $H^{orb}_{c2}(0)$ is 0.92(1) T for $\alpha$ = 0.69. Since the upper critical field values for both in-plane and out-of-plane orientations lie well below the Pauli limit, this signifies the role of the orbital effect.

In anisotropic superconductors, the relationship between the superconducting coherence length ($\xi$) and the upper critical field ($H_{c2}$) is quantitatively described by a well-established theoretical model \cite{palstra1988angular}, as follows: 
\begin{equation}
    H_{c2} = \frac{\phi_0}{2\pi\xi_{\perp c}^2}(cos^2\theta+\epsilon^2sin^2\theta)^{-1/2}.
    \label{eqn4: coherence}
\end{equation} 
Here $\phi_{0}$ = h/2e is the magnetic flux quanta, with a value of 2.07$\times$10$^{-15}$ T$m^{2}$. The parameter $\epsilon$ is the anisotropy ratio between the two coherence lengths along the in-plane and out-of-plane directions, defined as $\epsilon$ = $\xi_{\parallel c}$/$\xi_{\perp c}$. $\theta$ refers to the angle between the direction of the applied magnetic field and the unit vector normal to the superconducting layers. By evaluating the general expression for the angular dependence of the upper critical field [\equref{eqn4: coherence}] in the limiting cases of $\theta = 0^\circ$ and $90^\circ$, we arrive at simplified formulas for the coherence lengths in the parallel and perpendicular directions relative to the field. Specifically, for $\theta = 0^\circ$, the upper critical field is given by $H_{c2}^{\parallel}(0) = \frac{\phi_0}{2\pi \xi_{\perp}^2}$, and for $\theta = 90^\circ$, it is $H_{c2}^{\perp}(0) = \frac{\phi_0}{2\pi \xi_{\parallel} \xi_{\perp}}$. So, using these relations, we calculate the coherence lengths to be $\xi_{\parallel c}$ = 8.82(1) nm and $\xi_{\perp c}$ = 23.2(4) nm, respectively.  

A series of usual theoretical relations are applied to acquire the GL-penetration length ($\lambda$) and the GL parameter ($\kappa$): $H_{c2}^\perp$(0)/$H_{c1}^\perp$(0) = 2$\kappa_{\perp c}^2/ln \kappa_{\perp c}$, $\kappa_{\perp c}$ = [$\lambda_{\perp c}$(0)$\lambda_{\parallel c}$(0)/$\xi_{\perp c}$(0)$\xi_{\parallel c}$(0)]$^{1/2}$, and $\kappa_{\parallel c}$ = $\lambda_{\perp c}$(0)/$\xi_{\perp c}$(0). The complete set of extracted superconducting parameters is listed in \tableref{tbl1}. The calculated values of $\kappa$ for both orientations significantly exceed the critical threshold of $1/\sqrt{2}$, thereby unambiguously confirming that ZrTe$_{2.8}$ belongs to the class of strong type-II superconductors. Moreover, the thermodynamic critical field ($H_c$), which serves as a measure of the superconducting condensation energy, was found to be around 0.024(8) T, with $H_c(0) = H_{c1}^\perp(0)\sqrt{2}\kappa_{\perp c}/ln \kappa_{\perp c}$. \tableref{tbl2} lists the superconducting parameters of other Zr-based superconductors.

\begin{table}[b]
\caption{Superconducting parameters of ZrTe$_{2.8}$ with other Zr-based compounds.}
\label{tbl2}
\setlength{\tabcolsep}{7pt}
\begin{center}
\renewcommand{\arraystretch}{1.2}
\begin{tabular}[b]{lcccc}\hline
Compounds& $T_c$ (K)& $H\parallel c$ (T)& $H\perp c$ (T)& $\Gamma$ \\
\hline
ZrTe$_{3-x}$ & 3.59& 0.61& 1.61& 0.38 \\
ZrTe$_{3-x}$ \\ nanoplates & 3.4& 0.5& 2.5& 0.2 \\
Cu$_{0.05}$ZrTe$_3$ & 3.8&  1.3& 3.2& 0.41 \\ 
Ni$_{0.05}$ZrTe$_3$ & 3.1& 0.48& 1.23& 0.39 \\
\hline
\end{tabular}
\end{center}
\end{table}

\subsection{Specific heat}
\begin{figure}
\includegraphics[width=.95\columnwidth]{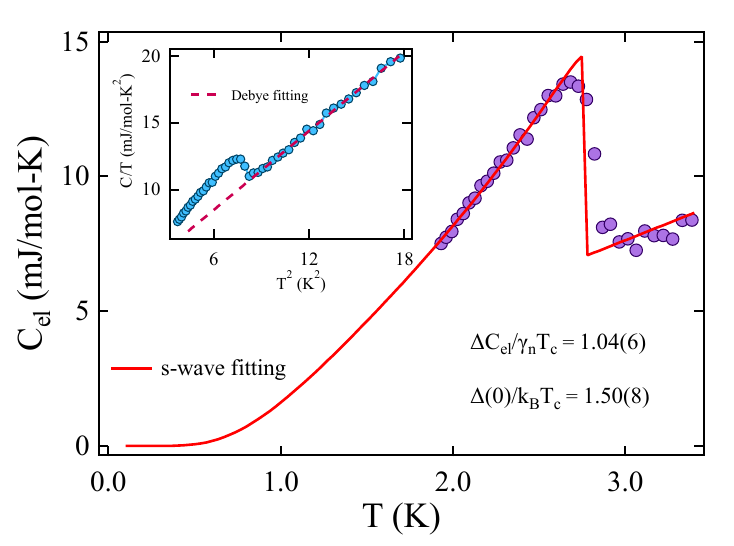}
\caption{\label{fig3} Electronic specific heat is fitted with an isotropic s-wave model, supporting conventional superconducting behavior. Inset: Debye-Sommerfeld model fit to the normal-state specific heat data.}
\end{figure}

The occurrence of the superconducting transition is further supported by a pronounced discontinuity in the zero-field-specific heat data, which affirms the presence of bulk superconductivity in this layered material. The system undergoes a second-order phase transition at 2.7 K, which closely matches the values $T_c$ identified in the resistivity and magnetization measurements. To analyze the behavior of the normal-state, the specific heat data above the transition were fitted using the Debye-Sommerfeld model, expressed as $C(T) = \gamma_{n}T + \beta_{3}T^{3}$, where the first term represents the electronic contribution to the heat capacity, governed by the Sommerfeld coefficient $\gamma_n$, and the second term accounts for the phononic (lattice) contribution at low temperatures through the Debye constant $\beta_3$. This fitting, illustrated in the inset of \figref{fig3}, yields the parameters $\gamma_{n} = 2.54(3)$ mJ mol$^{-1}$ K$^{-2}$ and $\beta_{3} = 0.98(8)$ mJ mol$^{-1}$ K$^{-4}$. The extracted value $\beta_3$ was then used to calculate the Debye temperature $\theta_D$, which provides valuable information about the phonon spectrum and vibrational dynamics of the crystal lattice. The Debye temperature is determined using the following relation:
\begin{equation}
\theta_{D} = \left(\frac{12\pi^{4} R N}{5 \beta_{3}}\right)^{\frac{1}{3}},
\label{eqn8:DebyeTemperature}
\end{equation}
where $R$ = 8.314 J mol$^{-1}$ K$^{-1}$ is the universal gas constant, and 
$N$ = 4 corresponds to the number of atoms per formula unit. The obtained Debye temperature, $\theta_{D}$, is found to be approximately 195.98(2) K, which is slightly lower than the value obtained from the fit of BG to the resistivity in the normal state. This discrepancy likely arises from the use of different temperature ranges in the respective analyzes. 

In BCS-type superconductors, the electronic density of states at the Fermi level, $N(E_{F})$, is a fundamental parameter that significantly influences the superconducting pairing mechanism. It governs the availability of electronic states that participate in the formation of Cooper pairs. This quantity can be estimated by incorporating $\gamma_{n}$ in $\gamma_{n}$ = $\left(\frac{\pi^{2} k_{B}^{2}}{3}\right) N(E_{F})$, where $k_{B}$ = 1.38 $\times$ 10$^{-23}$ J K$^{-1}$. The resulting value of $N(E_{F})$ is 1.80(6) states eV$^{-1}$f.u.$^{-1}$. The modified McMillan equation \cite{mcmillan1968transition} is used to evaluate the electron-phonon coupling constant, $\lambda_{e-ph}$, which measures the strength of the interaction between electrons and lattice vibrations (phonons). It is written as follows:
\begin{equation}
\lambda_{e-ph} = \frac{1.04+\mu^{*}\mathrm{ln}(\theta_{D}/1.45T_{c})}{(1-0.62\mu^{*})\mathrm{ln}(\theta_{D}/1.45T_{c})-1.04 };
\label{eqn9:Lambda}
\end{equation}
where $\mu^{*}$ represents the Coulomb pseudopotential, which accounts for electron-electron repulsion and is generally assumed to be 0.13 for transition metal-based superconductors. Applying this formalism with $T_{c}$ = 3.59(4) K and $\theta_{D}$ = 195.46(1) K, the result $\lambda_{e-ph}$ is 0.61(9). This value indicates that the compound studied belongs to the category of weakly coupled BCS superconductors. 

\begin{figure*}
\includegraphics[width=1.5\columnwidth]{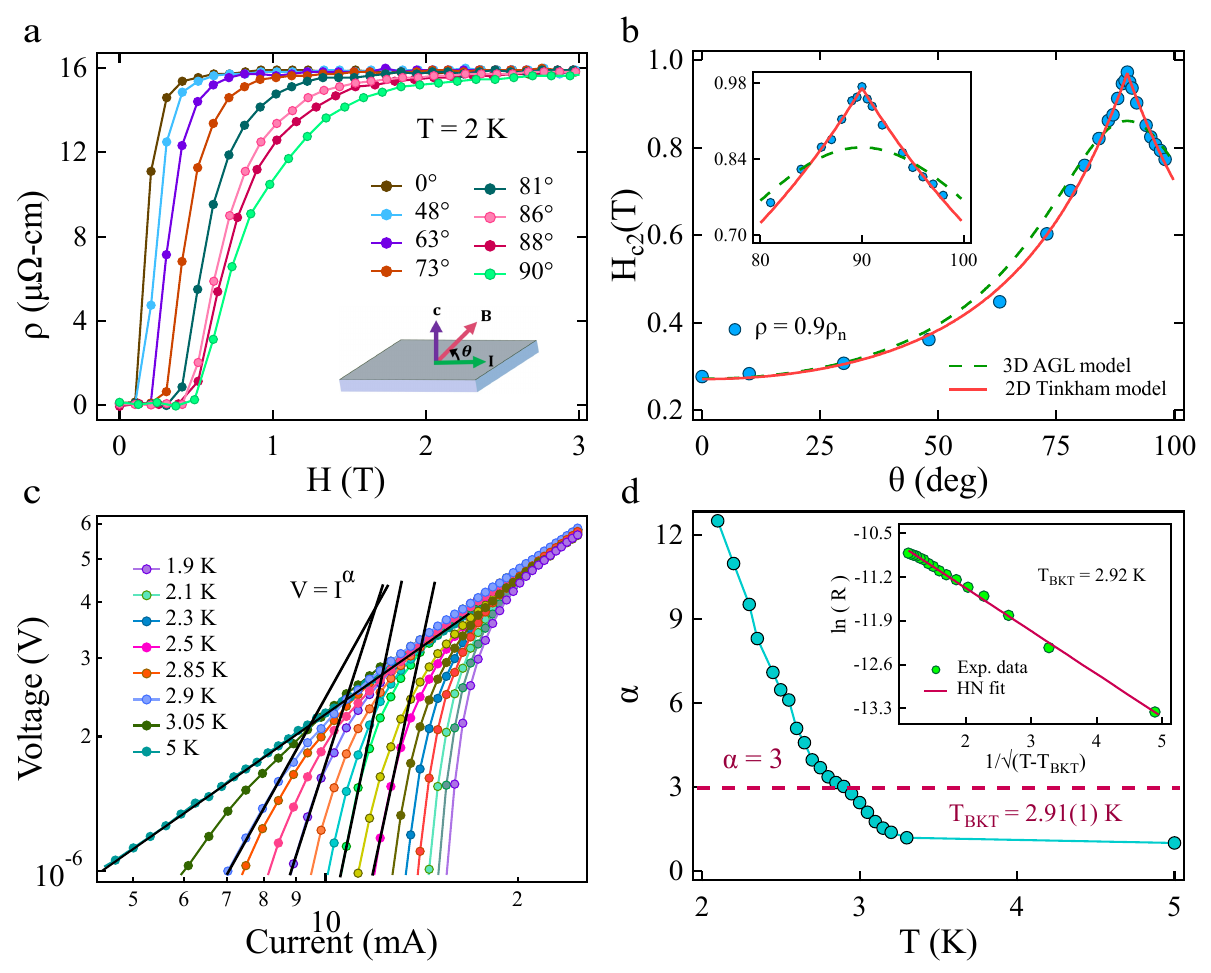}
\caption {\label{fig4} (a) Field-dependent electrical resistivity with different angles supports an anisotropic nature of this compound. Inset: Schematic of a tilt experimental setup. (b) Angle-dependent upper critical fields $\rho = 0.9\rho_n$ are best portrayed by the two-dimensional Tinkham model over the three-dimensional anisotropic GL model. Inset: expanded view of the fitting near 90$^{\circ}$ angle. (c) The temperature-dependent current-voltage ($V$-$I$) curves at zero magnetic field are plotted on a logarithmic scale. (d) The exponent ($\alpha$) values from power-law fitting correspond to the BKT transition temperature at 2.91(1) K. The slope of $\alpha (T)$ = 3 is marked by the horizontal dotted line. Inset: The Halperin-Nelson fitting is also incorporated to calculate the BKT transition temperature, indicating $T_{BKT}$ at 2.92 K.}
\end{figure*}

Important information about the underlying pairing mechanism of the superconducting state may be gained from the temperature dependence of the electron-specific heat data. To isolate the electronic contribution, the phononic part of the zero-field specific heat is deducted using the formula $C_{el} = C - \beta_{3}T^{3}$, where $\beta_{3}$ is the lattice contribution extracted from the normal-state fit. The resulting data $C_{el}$ are analyzed using a single-gap s-wave BCS model \cite{padamsee1973quasiparticle}, as shown in \figref{fig3}. Entropy $S$ can be assessed using this paradigm, as
\begin{equation}
\begin{split}
\frac{S}{\gamma_{n} T_{c}}= -\frac{6}{\pi^{2}} \left(\frac{\Delta(0)}{k_{B} T_{C}}\right) &\int_{0}^{\infty}[ fln(f)\\
&+(1-f)ln(1-f)] dy,
\end{split}
\label{Eq:swave}
\end{equation}
where, $f(\xi)$ = $[e^{\beta E(\xi)}+1]^{-1}$ is the Fermi-Dirac distribution function, where $\beta = 1/k_BT$ and $E(\xi) = \sqrt{\xi^2 + \Delta^2(t)}$ is the quasiparticle excitation energy. The variable $\xi$ denotes the energy of normal-state electrons measured relative to the Fermi level, while $\Delta(t)$ is the superconducting energy gap as a function of the reduced temperature $t = T/T_c$. For the purpose of integration, the variable substitution $y = \xi/\Delta(0)$ is applied. The temperature-dependent superconducting gap is described by the empirical formula: $\Delta(t) = tanh[1.82(1.018[(1/t)-1])^{0.51}]$, implementing the isotropic s-wave BCS approximation. The electronic specific heat is then related to the entropy via the thermodynamic relation $C_{el}(T) = t \frac{dS}{dt}$. By fitting the experimental data $C_{el}(T)$ using this formalism, the superconducting gap ratio $\Delta(0)/k_B T_c$ is determined to be approximately 1.50(8), which is in excellent agreement with the prediction of BCS for weak-coupling superconductors. Furthermore, the specific heat jump at $T_c$, denoted as $\Delta C_{el}/\gamma_n T_c$, is found to be nearly 1.04(6), further supporting the classification of this material as a conventional s-wave superconductor.

\subsection{Two-dimensional superconductivity}
\textbf{Angle-dependent upper critical field:} The angle-dependent upper critical field is a useful probe for identifying anisotropy and possible two-dimensional superconducting behavior. To investigate this, field-dependent resistivity measurements were performed at a stable temperature of 2 K for various angles, as shown in \figref{fig4}(a), where $\theta$ is defined as the angle between the applied magnetic field and the normal to the sample plane, with the current always along the $ab$ plane. The upper critical field at each angle was extracted using the criterion $\rho = 0.9\rho_n$, where $\rho_n$ is the resistivity in the normal state. As shown in \figref{fig4}(b), the angular dependence of $H_{c2}$ displays a pronounced cusp near $\theta = 90^\circ$, which becomes more apparent in the magnified view provided in the inset. Such a sharp angular feature is a hallmark of strong anisotropy and often suggests the presence of quasi-two-dimensional superconductivity. Two theoretical models are commonly considered to explain the angular dependence of $H_{c2}$. The anisotropic Ginzburg–Landau (AGL) model (\equref{eqnAGL}) describes the angular variation for three-dimensional superconductors, resulting in an ellipsoidal form of $H_{c2}(\theta)$. In contrast, Tinkham’s model (\equref{eqn2DT}) captures the behavior in two-dimensional thin-film superconductors and predicts a characteristic cusp at $\theta = 90^\circ$. Comparing the experimental data with these models allows us to assess the dimensionality and anisotropy of the superconducting state in this system.
\begin{equation}\label{eqnAGL}
    \left(\frac{H_{c2}(\theta,T) \sin{\theta}}{H_{c2}^{\perp}}\right)^2 + \left(\frac{H_{c2}( \theta,T) \cos{\theta}}{H_{c2}^{||}}\right)^2 = 1
\end{equation}

\begin{equation}\label{eqn2DT}
    \left(\frac{H_{c2}(\theta,T) \sin{\theta}}{H_{c2}^{\perp}}\right)^2 + \left\arrowvert\frac{H_{c2}( \theta,T) \cos{\theta}}{H_{c2}^{||}}\right\arrowvert = 1
\end{equation}

The experimental data are better described by the two-dimensional Tinkham model than by the three-dimensional AGL model \cite{patra2024planar}, indicating a more pronounced two-dimensional nature of superconductivity. Collectively, these results suggest that the superconductivity in ZrTe$_{2.8}$ exhibits a quasi-two-dimensional behavior.

\textbf{Berezinskii-Kosterlitz-Thouless (BKT) transition:} In addition to angle-dependent upper critical field measurements, a powerful complementary approach to probe the two-dimensional nature of superconductivity is through the identification of a Berezinskii-Kosterlitz-Thouless (BKT) transition in the transport behavior of ZrTe$_{3-x}$. In contrast to three-dimensional superconductors, which can support true long-range phase coherence, two-dimensional superconductors are limited to exhibiting quasi-long-range order due to enhanced thermal fluctuations. The BKT transition emerges as the system transitions from a quasi-long-range ordered vortex–antivortex phase to a fully disordered state with increasing temperature \cite{takiguchi2024berezinskii}. The temperature-dependent current-voltage characteristics ($V$-$I$) (as plotted in \figref{fig4}(c)) are used to extract the BKT transition temperature, $T_{BKT}$. Below $T_c$, the application of high currents can break the vortex-antivortex pairs, resulting in a non-ohmic response. In this regime, the $V$-$I$ characteristic nicely follows a power-law dependence (\equref{eqnV-I}), where the exponent is directly proportional to the superfluid density ($J_s$). The equilibrium density of the free vortices $n_v$(I) scales as a power of the applied current, which is the typical origin of the behavior mentioned above. Consequently, the voltage response behaves as $V \propto$ $n_v(I)$$I$, capturing the nonlinear dynamics of the vortex motion in the quasi-2D superconducting state \cite{PhysRevB.100.064506}. In the ideal scenario of a BKT transition, it is predicted that the superfluid density $J_s$ exhibits a universal and discontinuous jump at the critical temperature where it intersects the BKT line. When this behavior is substituted for \equref{eqnV-I}, it implies that the voltage–current exponent should also show an abrupt change at the transition point, which serves as a hallmark of the BKT mechanism.
\begin{equation}
    V \propto I^{\alpha(T)},  
    \quad \alpha(T) = 1 + \pi \frac{J_s(T)}{T} .
    \label{eqnV-I}
\end{equation}
\begin{equation}
    J_s(T^-_{BKT}) = \frac{2}{\pi T_{BKT}},  
     \quad J_s(T^+_{BKT}) = 0.
    \label{eqnBKT}
\end{equation}
\begin{equation}
    \alpha(T^-_{BKT}) = 3,  
    \quad \alpha(T^+_{BKT}) = 1.
    \label{eqnalpha}
\end{equation}

By fitting the power-law dependence described by \equref{eqnV-I}, we extracted the temperature-dependent exponent $\alpha(T)$, as shown in \figref{fig4}(d). The extracted $\alpha$ increases with decreasing temperature and approaches a value of 3 near 2.91(1) K, which is identified as $T_{BKT}$.

Additionally, to check the consistency of the BKT behavior, another theoretical approach proposed by Halperin and Nelson is used. According to the Halperin-Nelson equation (HN), in the temperature range just above $T_{BKT}$, the resistance follows an exponential dependence given by \equref{eqnHN}.
\begin{equation}
    R = R_0exp[\frac{-b}{(T-T_{BKT})^{1/2}}].
    \label{eqnHN}
\end{equation}

Here, $R_0$ is a material-specific constant, and $b$ tells about the strength of the vortex-antivortex interaction. This expression is valid within a narrow temperature window above $T_{BKT}$, where superconductivity is suppressed by phase fluctuations induced by thermal unbinding of vortex-antivortex pairs. The plot of $lnR$ versus $1/(T-T_{BKT})^{1/2}$ is linearly fitted to estimate $T_{BKT}$ (Inset of \figref{fig4}(d)). Yielded value of $T_{BKT}$ is 2.92 K, which aligns well with the estimate obtained from the non-linear $I$-$V$ power-law analysis.

The agreement between these two independent methods, nonlinear scaling $V$-$I$ and Halperin–Nelson fitting, strongly supports the presence of a Berezinskii-Kosterlitz-Thouless transition, reinforcing the quasi-two-dimensional nature of superconductivity in this system.

\section{Conclusion}
In summary, we have synthesized single crystals of ZrTe$_{3-x}$ using the isothermal chemical vapor transport method. Magnetization, specific heat, and transport measurements were conducted on a crystal, revealing bulk anisotropic type-II superconductivity at a temperature of 3.59(4) \si{K}. The observation of an isotropic s-wave superconducting gap further classifies the compound as a weakly coupled conventional superconductor. More importantly, the presence of a quasi-2D nature is confirmed by angle-dependent upper critical field measurements conforming to the 2D Tinkham model and observation of the BKT transition at 2.91(1) \si{K}. Our results suggest that the introduction of Te vacancies can be used as an effective tuning parameter to modulate the balance between superconductivity and the charge density wave transition. This approach not only uncovers the relation between these competing ground states, but also offers an opportunity to explore numerous low-dimensional quantum phases in bulk materials of group IVB TMTCs, considerably broadening the new pathway for pursuing two-dimensional superconductivity in layered materials.

\section{Acknowledgments} 
P.~M. acknowledges the funding agency DST-INSPIRE, Government of India, for providing the SRF fellowship. R.~P.~S. acknowledges the Science and Engineering Research Board, Government of India, for the Core Research Grant CRG/2023/000817.

\nocite{*}
\bibliographystyle{revtex}
\bibliography{Library}

@article{PhysRevB.100.064506,
  title = {Nonlinear $I\text{\ensuremath{-}}V$ characteristics of two-dimensional superconductors: Berezinskii-Kosterlitz-Thouless physics versus inhomogeneity},
  author = {Venditti, G. and Biscaras, J. and Hurand, S. and Bergeal, N. and Lesueur, J. and Dogra, A. and Budhani, R. C. and Mondal, Mintu and Jesudasan, John and Raychaudhuri, Pratap and Caprara, S. and Benfatto, L.},
  journal = {Phys. Rev. B},
  volume = {100},
  issue = {6},
  pages = {064506},
  numpages = {10},
  year = {2019},
  month = {Aug},
  publisher = {American Physical Society},
  doi = {10.1103/PhysRevB.100.064506},
  url = {https://link.aps.org/doi/10.1103/PhysRevB.100.06450}
}

@article{Chandrasekhar1962pauli,
  author       = {Chandrasekhar, B S},
  title        = {A note on the maximum critical field of high-field superconductors},
  doi          = {10.1063/1.1777362},
  url          = {https://www.osti.gov/biblio/4734493},
  journal      = {Appl. Phys. Lett.},
  volume       = {1},
  pages        = {7--8},
  place        = {Country unknown/Code not available},
  year         = {1962}
}

@article{Clogston1962pauli,
  title = {Upper Limit for the Critical Field in Hard Superconductors},
  author = {Clogston, A. M.},
  journal = {Phys. Rev. Lett.},
  volume = {9},
  issue = {6},
  pages = {266--267},
  year = {1962},
  month = {Sep},
  publisher = {American Physical Society},
  doi = {10.1103/PhysRevLett.9.266},
  url = {https://link.aps.org/doi/10.1103/PhysRevLett.9.266}
}

@article{WHH1966orbital,
  title = {Temperature and Purity Dependence of the Superconducting Critical Field, ${H}_{c2}$. III. Electron Spin and Spin-Orbit Effects},
  author = {Werthamer, N. R. and Helfand, E. and Hohenberg, P. C.},
  journal = {Phys. Rev.},
  volume = {147},
  issue = {1},
  pages = {295--302},
  year = {1966},
  month = {Jul},
  publisher = {American Physical Society},
  doi = {10.1103/PhysRev.147.295},
  url = {https://link.aps.org/doi/10.1103/PhysRev.147.295}
}

@article{Helfand1966orbital,
  title = {Temperature and Purity Dependence of the Superconducting Critical Field, ${H}_{c2}$. II},
  author = {Helfand, E. and Werthamer, N. R.},
  journal = {Phys. Rev.},
  volume = {147},
  issue = {1},
  pages = {288--294},
  numpages = {0},
  year = {1966},
  month = {Jul},
  publisher = {American Physical Society},
  doi = {10.1103/PhysRev.147.288},
  url = {https://link.aps.org/doi/10.1103/PhysRev.147.288}
}

@article{mcmillan1968transition,
  title = {Transition Temperature of Strong-Coupled Superconductors},
  author = {McMillan, W. L.},
  journal = {Phys. Rev.},
  volume = {167},
  issue = {2},
  pages = {331--344},
  year = {1968},
  month = {Mar},
  publisher = {American Physical Society},
  doi = {10.1103/PhysRev.167.331},
  url = {https://link.aps.org/doi/10.1103/PhysRev.167.331}
}

@article{padamsee1973quasiparticle,
  title={Quasiparticle phenomenology for thermodynamics of strong-coupling superconductors},
  author={Padamsee, H and Neighbor, JE and Shiffman, CA},
  journal={J. Low Temp. Phys.},
  volume={12},
  pages={387--411},
  year={1973},
  doi = {https://doi.org/10.1007/BF00654872},
  publisher={Springer}
}

@article{fullprof,
title = {Recent advances in magnetic structure determination by neutron powder diffraction},
journal = {Physica B: Condens. Matter},
volume = {192},
number = {1},
pages = {55-69},
year = {1993},
issn = {0921-4526},
doi = {https://doi.org/10.1016/0921-4526(93)90108-I},
author = {Juan Rodríguez-Carvajal}
}

@misc{Supp,
    title = {See supplemental material for additional details on powder {XRD} data, {EDAX} data, Magnetization, Resistivity, and {Uemura} plot, which includes Refs. [46-58]}
}

@article{momma2011vesta,
  title={VESTA 3 for three-dimensional visualization of crystal, volumetric and morphology data},
  author={Momma, Koichi and Izumi, Fujio},
  journal={J. Appl. Cryst.},
  volume={44},
  number={6},
  pages={1272--1276},
  year={2011},
  doi={https://doi.org/10.1107/S0021889811038970},
  publisher={International Union of Crystallography}
}

@article{lian2023interplay,
  title={Interplay of charge ordering and superconductivity in two-dimensional 2$H$ group V transition-metal dichalcogenides},
  author={Lian, Chao-Sheng},
  journal={Phys. Rev. B},
  volume={107},
  number={4},
  pages={045431},
  year={2023},
  publisher={APS},
  doi={https://doi.org/10.1103/PhysRevB.107.045431}
}

@article{palstra1988angular,
  title={Angular dependence of the upper critical field of Bi$_{2.2}$Sr$_2$Ca$_{0.8}$Cu$_2$O$_{8+\delta}$},
  author={Palstra, TTM and Batlogg, Bertram and Schneemeyer, LF and Van Dover, RB and Waszczak, Joseph V},
  journal={Phys. Rev. B},
  volume={38},
  number={7},
  pages={5102},
  year={1988},
  publisher={APS},
  doi={https://doi.org/10.1103/PhysRevB.38.5102}
}

@article{takiguchi2024berezinskii,
  title={Berezinskii-Kosterlitz-Thouless transition in rhenium nitride films},
  author={Takiguchi, Kosuke and Krockenberger, Yoshiharu and Taniyasu, Yoshitaka and Yamamoto, Hideki},
  journal={Phys. Rev. B},
  volume={110},
  number={2},
  pages={024516},
  year={2024},
  publisher={APS},
  doi={https://doi.org/10.1103/PhysRevB.110.024516}
}

@article{PhysRevB.91.155124,
  title = {Structural contributions to the pressure-tuned charge-density-wave to superconductor transition in ${\mathrm{ZrTe}}_{3}$: Raman scattering studies},
  author = {Gleason, S. L. and Gim, Y. and Byrum, T. and Kogar, A. and Abbamonte, P. and Fradkin, E. and MacDougall, G. J. and Van Harlingen, D. J. and Zhu, Xiangde and Petrovic, C. and Cooper, S. L.},
  journal = {Phys. Rev. B},
  volume = {91},
  issue = {15},
  pages = {155124},
  numpages = {6},
  year = {2015},
  month = {Apr},
  publisher = {American Physical Society},
  doi = {10.1103/PhysRevB.91.155124},
  url = {https://link.aps.org/doi/10.1103/PhysRevB.91.155124}
}

@article{PhysRevB.80.075423,
  title = {Splitting in the Fermi surface of ${\text{ZrTe}}_{3}$: A surface charge density wave system},
  author = {Hoesch, Moritz and Cui, Xiaoyu and Shimada, Kenya and Battaglia, Corsin and Fujimori, Shin-ichi and Berger, Helmuth},
  journal = {Phys. Rev. B},
  volume = {80},
  issue = {7},
  pages = {075423},
  numpages = {8},
  year = {2009},
  month = {Aug},
  publisher = {American Physical Society},
  doi = {10.1103/PhysRevB.80.075423},
  url = {https://link.aps.org/doi/10.1103/PhysRevB.80.075423}
}

@article{gu2018pressure,
  title={Pressure-induced enhancement in the superconductivity of ZrTe$_3$},
  author={Gu, Kemin and Susilo, Resta A and Ke, Feng and Deng, Wen and Wang, Yanju and Zhang, Lingkong and Xiao, Hong and Chen, Bin},
  journal={J. Phys. : Condens. Matter},
  volume={30},
  number={38},
  pages={385701},
  year={2018},
  publisher={IOP Publishing},
  doi={10.1088/1361-648X/aada53}
}

@article{Lei_2011,
doi = {10.1209/0295-5075/95/17011},
url = {https://dx.doi.org/10.1209/0295-5075/95/17011},
year = {2011},
month = {jun},
publisher = {},
volume = {95},
number = {1},
pages = {17011},
author = {Lei, Hechang and Zhu, Xiangde and Petrovic, C.},
title = {Raising Tc in charge density wave superconductor ZrTe$_3$ by Ni intercalation},
journal = {Europhys. Lett.},
}

@article{PhysRevLett.106.246404,
  title = {Coexistence of Bulk Superconductivity and Charge Density Wave in ${\mathrm{Cu}}_{x}{\mathrm{ZrTe}}_{3}$},
  author = {Zhu, Xiangde and Lei, Hechang and Petrovic, C.},
  journal = {Phys. Rev. Lett.},
  volume = {106},
  issue = {24},
  pages = {246404},
  numpages = {4},
  year = {2011},
  month = {Jun},
  publisher = {American Physical Society},
  doi = {10.1103/PhysRevLett.106.246404},
  url = {https://link.aps.org/doi/10.1103/PhysRevLett.106.246404}
}

@article{PhysRevB.71.132508,
  title = {Pressure effect on competition between charge density wave and superconductivity in ${\mathrm{ZrTe}}_{3}$: Appearance of pressure-induced reentrant superconductivity},
  author = {Yomo, R. and Yamaya, K. and Abliz, M. and Hedo, M. and Uwatoko, Y.},
  journal = {Phys. Rev. B},
  volume = {71},
  issue = {13},
  pages = {132508},
  numpages = {4},
  year = {2005},
  month = {Apr},
  publisher = {American Physical Society},
  doi = {10.1103/PhysRevB.71.132508},
  url = {https://link.aps.org/doi/10.1103/PhysRevB.71.132508}
}

@article{PhysRevB.87.024508,
  title = {Disorder-induced bulk superconductivity in ZrTe${}_{3}$ single crystals via growth control},
  author = {Zhu, Xiyu and Lv, Bing and Wei, Fengyan and Xue, Yuyi and Lorenz, Bernd and Deng, Liangzi and Sun, Yanyi and Chu, Ching-Wu},
  journal = {Phys. Rev. B},
  volume = {87},
  issue = {2},
  pages = {024508},
  numpages = {6},
  year = {2013},
  month = {Jan},
  publisher = {American Physical Society},
  doi = {10.1103/PhysRevB.87.024508},
  url = {https://link.aps.org/doi/10.1103/PhysRevB.87.024508}
}

@article{wang2023superconductivity,
  title={Superconductivity in single-crystalline ZrTe$_{3-x}$ (x $\leq$ 0.5) nanoplates},
  author={Wang, Jie and Wu, Min and Zhen, Weili and Li, Tian and Li, Yun and Zhu, Xiangde and Ning, Wei and Tian, Mingliang},
  journal={Nanoscale Adv.},
  volume={5},
  number={2},
  pages={479--484},
  year={2023},
  publisher={Royal Society of Chemistry},
  doi={10.1039/D2NA00628F}
}

@article{patra2020anisotropic,
  title={Anisotropic quasi-one-dimensional layered transition-metal trichalcogenides: synthesis, properties and applications},
  author={Patra, Abhinandan and Rout, Chandra Sekhar},
  journal={RSC adv.},
  volume={10},
  number={60},
  pages={36413--36438},
  year={2020},
  publisher={Royal Society of Chemistry},
  doi={10.1039/D0RA07160A}
}

@article{patra2024planar,
  title={Planar Hall Effect and Quasi-2D Anisotropic Superconductivity in Topological Candidate 1$T$-NbSeTe},
  author={Patra, Chandan and Agarwal, Tarushi and Srivastava, Shashank and Chowdhury, Rajeswari Roy and Saravanan, MP and Singh, Ravi Prakash},
  journal={Adv. Quantum Technol.},
  volume={7},
  number={2},
  pages={2300448},
  year={2024},
  publisher={Wiley Online Library},
  doi={https://doi.org/10.1002/qute.202300448}
}

@article{PhysRevB.85.184513,
  title = {Mixed bulk-filament nature in superconductivity of the charge-density-wave conductor ZrTe${}_{3}$},
  author = {Yamaya, Kazuhiko and Takayanagi, Shigeru and Tanda, Satoshi},
  journal = {Phys. Rev. B},
  volume = {85},
  issue = {18},
  pages = {184513},
  numpages = {6},
  year = {2012},
  month = {May},
  publisher = {American Physical Society},
  doi = {10.1103/PhysRevB.85.184513},
  url = {https://link.aps.org/doi/10.1103/PhysRevB.85.184513}
}

@article{doi:10.1021/acsnano.9b02196,
author = {Wang, Huichao and Chan, Cheuk Ho and Suen, Chun Hung and Lau, Shu Ping and Dai, Ji-Yan},
title = {Magnetotransport Properties of Layered Topological Material ZrTe$_2$ Thin Film},
journal = {ACS Nano},
volume = {13},
number = {5},
pages = {6008-6016},
year = {2019},
doi = {10.1021/acsnano.9b02196},
note ={PMID: 31013050},
URL ={https://doi.org/10.1021/acsnano.9b02196}
}

@article{wang2022magnetotransport,
  title={Magnetotransport due to conductivity fluctuations in non-magnetic ZrTe$_2$ nanoplates},
  author={Wang, Jie and Wang, Yihao and Wu, Min and Li, Junbo and Miao, Shaopeng and Hou, Qingyi and Li, Yun and Zhou, Jianhui and Zhu, Xiangde and Xiong, Yimin and others},
  journal={Appl. Phys. Lett.},
  volume={120},
  number={16},
  year={2022},
  publisher={AIP Publishing},
  url={https://doi.org/10.1063/5.0083154}
}

@article{PhysRevB.94.165201,
  title = {Coexistence of Weyl fermion and massless triply degenerate nodal points},
  author = {Weng, Hongming and Fang, Chen and Fang, Zhong and Dai, Xi},
  journal = {Phys. Rev. B},
  volume = {94},
  issue = {16},
  pages = {165201},
  numpages = {7},
  year = {2016},
  month = {Oct},
  publisher = {American Physical Society},
  doi = {10.1103/PhysRevB.94.165201},
  url = {https://link.aps.org/doi/10.1103/PhysRevB.94.165201}
}

@article{bouhon2020non,
  title={Non-Abelian reciprocal braiding of Weyl points and its manifestation in ZrTe},
  author={Bouhon, Adrien and Wu, QuanSheng and Slager, Robert-Jan and Weng, Hongming and Yazyev, Oleg V and Bzdu{\v{s}}ek, Tom{\'a}{\v{s}}},
  journal={Nat. Phys.},
  volume={16},
  number={11},
  pages={1137--1143},
  year={2020},
  publisher={Nature Publishing Group UK London},
  doi={https://doi.org/10.1038/s41567-020-0967-9}
}

@article{doi:10.1126/science.1223532,
author = {G. Ghiringhelli and M. Le Tacon and M. Minola and S. Blanco-Canosa and C. Mazzoli and N. B. Brookes and G. M. De Luca and A. Frano and D. G. Hawthorn and F. He  and T. Loew  and M. Moretti Sala  and D. C. Peets  and M. Salluzzo  and E. Schierle  and R. Sutarto  and G. A. Sawatzky  and E. Weschke  and B. Keimer  and L. Braicovich},
title = {Long-Range Incommensurate Charge Fluctuations in (Y, Nd)Ba$_2$Cu$_3$O$_{6+x}$},
journal = {Science},
volume = {337},
number = {6096},
pages = {821-825},
year = {2012},
doi = {10.1126/science.1223532}
}

@article{PhysRevB.89.224513,
  title = {Charge density wave fluctuations in ${\text{La}}_{2\ensuremath{-}x}$${\text{Sr}}_{x}$${\text{CuO}}_{4}$ and their competition with superconductivity},
  author = {Croft, T. P. and Lester, C. and Senn, M. S. and Bombardi, A. and Hayden, S. M.},
  journal = {Phys. Rev. B},
  volume = {89},
  issue = {22},
  pages = {224513},
  numpages = {8},
  year = {2014},
  month = {Jun},
  publisher = {American Physical Society},
  doi = {10.1103/PhysRevB.89.224513},
  url = {https://link.aps.org/doi/10.1103/PhysRevB.89.224513}
}

@article{doi:10.1143/JPSJ.79.123710,
author = {Kudo, Kazutaka and Nishikubo, Yoshihiro and Nohara, Minoru},
title = {Coexistence of Superconductivity and Charge Density Wave in SrPt$_2$As$_2$},
journal = {J. Phys. Soc. Japan},
volume = {79},
number = {12},
pages = {123710},
year = {2010},
doi = {10.1143/JPSJ.79.123710}
}

@article{RevModPhys.60.1129,
  title = {The dynamics of charge-density waves},
  author = {Gr\"uner, G.},
  journal = {Rev. Mod. Phys.},
  volume = {60},
  issue = {4},
  pages = {1129--1181},
  numpages = {0},
  year = {1988},
  month = {Oct},
  publisher = {American Physical Society},
  doi = {10.1103/RevModPhys.60.1129},
  url = {https://link.aps.org/doi/10.1103/RevModPhys.60.1129}
}

@article{petrovic2016disorder,
  title={A disorder-enhanced quasi-one-dimensional superconductor},
  author={Petrovi{\'c}, Alexander Paul and Ansermet, Diane and Chernyshov, Dmitry and Hoesch, M and Salloum, Diana and Gougeon, Patrick and Potel, Michel and Boeri, Lilia and Panagopoulos, Christos},
  journal={Nat. commun.},
  volume={7},
  number={1},
  pages={12262},
  year={2016},
  publisher={Nature Publishing Group UK London},
  doi = {https://doi.org/10.1038/ncomms12262}
}

@article{PhysRevLett.122.017601,
  title = {Disorder Quenching of the Charge Density Wave in ${\mathrm{ZrTe}}_{3}$},
  author = {Hoesch, Moritz and Gannon, Liam and Shimada, Kenya and Parrett, Benjamin J. and Watson, Matthew D. and Kim, Timur K. and Zhu, Xiangde and Petrovic, Cedomir},
  journal = {Phys. Rev. Lett.},
  volume = {122},
  issue = {1},
  pages = {017601},
  numpages = {5},
  year = {2019},
  month = {Jan},
  publisher = {American Physical Society},
  doi = {10.1103/PhysRevLett.122.017601}
}

@article{chang2012direct,
  title={Direct observation of competition between superconductivity and charge density wave order in YBa$_2$Cu$_3$O$_{6.67}$},
  author={Chang, J and Blackburn, Elizabeth and Holmes, AT and Christensen, Niels B and Larsen, Jacob and Mesot, J and Liang, Ruixing and Bonn, DA and Hardy, WN and Watenphul, A and others},
  journal={Nat. Phys.},
  volume={8},
  number={12},
  pages={871--876},
  year={2012},
  publisher={Nature Publishing Group UK London}, 
  doi = {https://doi.org/10.1038/nphys2456}
}

@article{le2014inelastic,
  title={Inelastic X-ray scattering in YBa$_2$Cu$_3$O$_{6.6}$ reveals giant phonon anomalies and elastic central peak due to charge-density-wave formation},
  author={Le Tacon, M and Bosak, A and Souliou, SM and Dellea, Greta and Loew, T and Heid, R and Bohnen, KP and Ghiringhelli, G and Krisch, M and Keimer, B},
  journal={Nat. Phys.},
  volume={10},
  number={1},
  pages={52--58},
  year={2014},
  publisher={Nature Publishing Group UK London},
  doi = {https://doi.org/10.1038/nphys2805}
}

@article{
doi:10.1126/sciadv.aau5096,
author = {Huichao Wang  and Haiwen Liu  and Yanan Li  and Yongjie Liu  and Junfeng Wang  and Jun Liu  and Ji-Yan Dai  and Yong Wang  and Liang Li  and Jiaqiang Yan  and David Mandrus  and X. C. Xie  and Jian Wang },
title = {Discovery of log-periodic oscillations in ultraquantum topological materials},
journal = {Sci. Adv.},
volume = {4},
number = {11},
pages = {eaau5096},
year = {2018},
doi = {10.1126/sciadv.aau5096}
}

@article{tang2019three,
  title={Three-dimensional quantum Hall effect and metal--insulator transition in ZrTe$_5$},
  author={Tang, Fangdong and Ren, Yafei and Wang, Peipei and Zhong, Ruidan and Schneeloch, John and Yang, Shengyuan A and Yang, Kun and Lee, Patrick A and Gu, Genda and Qiao, Zhenhua and Liyuan, Zhang},
  journal={Nat.},
  volume={569},
  number={7757},
  pages={537--541},
  year={2019},
  publisher={Nature Publishing Group UK London},
  doi={https://doi.org/10.1038/s41586-019-1180-9}
}

@article{zhu2016superconductivity,
  title={Superconductivity and charge density wave in ZrTe$_{3-x}$Se$_x$},
  author={Zhu, Xiangde and Ning, Wei and Li, Lijun and Ling, Langsheng and Zhang, Ranran and Zhang, Jinglei and Wang, Kefeng and Liu, Yu and Pi, Li and Ma, Yongchang and Du, Haifeng and Tian, Minglian and Sun, Yuping and Petrovic, Cedomir and Zhang, Yuheng},
  journal={Sci. Rep.},
  volume={6},
  number={1},
  pages={26974},
  year={2016},
  publisher={Nature Publishing Group UK London},
  doi={https://doi.org/10.1038/srep26974}
}

@article{yamaya2002effect,
  title={The effect of pressure on the charge-density wave and superconductivity in ZrTe$_3$},
  author={Yamaya, K and Yoneda, M and Yasuzuka, S and Okajima, Y and Tanda, S},
  journal={J. Phys.: Condens. Matter},
  volume={14},
  number={44},
  pages={10767},
  year={2002},
  publisher={IOP Publishing},
  doi={10.1088/0953-8984/14/44/374}
}

@article{DEFARIA2024175919,
title = {Superconductivity, antiferromagnetism, and charge density waves in ZrTe$_3$ intercalated with Terbium},
journal = {J. Alloys and Compd.},
volume = {1005},
pages = {175919},
year = {2024},
issn = {0925-8388},
doi = {https://doi.org/10.1016/j.jallcom.2024.175919},
author = {L.R. {de Faria} and F. Abud and L.E. Correa and L.M. Ishikura and M.S. {da Luz} and M.S. Torikachvili and A.J.S. Machado}
}

@article{ISHIKURA2026131145,
title = {Superconductivity and charge density waves in ruthenium intercalated ZrTe$_3$},
journal = {Phys. Lett. A},
volume = {565},
pages = {131145},
year = {2026},
issn = {0375-9601},
doi = {https://doi.org/10.1016/j.physleta.2025.131145},
author = {Larissa Miki Ishikura and Fabio Santos Alves Abud and Leandro Rodrigues {de Faria} and Kipras Mikalajunas and Brunno Borges Canelhas and Deusmaque Carneiro Ferreira and Mário Sérgio {da Luz} and Milton Starosta Torikachvili and Antonio Jefferson da Silva Machado}
}

@article{manna2025quasi,
  title={Quasi-two-dimensional superconductivity in 1$T$-Ti$_{1-x}$Ta$_x$Se$_2$},
  author={Manna, P and Sharma, S and Agarwal, T and Srivastava, S and Mishra, P and Singh, RP},
  journal={arXiv:2511.00605},
  year={2025},
  url={https://doi.org/10.48550/arXiv.2511.00605}
}

@article{agarwal2023quasi,
  title={Quasi-two-dimensional anisotropic superconductivity in Li-intercalated 2$H$-TaS$_2$},
  author={Agarwal, Tarushi and Patra, C and Kataria, A and Chowdhury, Rajeswari R and Singh, RP},
  journal={Phys. Rev. B},
  volume={107},
  number={17},
  pages={174509},
  year={2023},
  publisher={APS},
  doi={https://doi.org/10.1103/PhysRevB.107.174509}
}

@article{patra2024superconducting,
  title={Superconducting Properties of Layered Chalocogenides 1$T$-RhSeTe},
  author={Patra, Chandan and Agarwal, Tarushi and Arushi and Manna, Poulami and Bhatt, Neeraj and Singh, Ravi Shankar and Singh, Ravi Prakash},
  journal={Adv. Quantum Technol.},
  volume={7},
  number={12},
  pages={2400175},
  year={2024},
  publisher={Wiley Online Library},
  doi={ https://doi.org/10.1002/qute.202400175}
}

@article{Uemura,
  title = {Universal Correlations between ${T}_{c}$ and $\frac{{n}_{s}}{{m}^{*}}$ (Carrier Density over Effective Mass) in High-${T}_{c}$ Cuprate Superconductors},
  author = {Uemura, Y. J. and Luke, G. M. and Sternlieb, B. J. and Brewer, J. H. and Carolan, J. F. and Hardy, W. N. and Kadono, R. and Kempton, J. R. and Kiefl, R. F. and Kreitzman, S. R. and Mulhern, P. and Riseman, T. M. and Williams, D. Ll. and Yang, B. X. and Uchida, S. and Takagi, H. and Gopalakrishnan, J. and Sleight, A. W. and Subramanian, M. A. and Chien, C. L. and Cieplak, M. Z. and Xiao, Gang and Lee, V. Y. and Statt, B. W. and Stronach, C. E. and Kossler, W. J. and Yu, X. H.},
  journal = {Phys. Rev. Lett.},
  volume = {62},
  issue = {19},
  pages = {2317--2320},
  numpages = {0},
  year = {1989},
  month = {May},
  publisher = {American Physical Society},
  doi = {10.1103/PhysRevLett.62.2317},
  url = {https://link.aps.org/doi/10.1103/PhysRevLett.62.2317}
}

@article{wang2017upper,
  title={The upper critical field and its anisotropy in (Li$_{1-x}$Fe$_x$)OHFe$_{1-y}$Se},
  author={Wang, Zhaosheng and Yuan, Jie and Wosnitza, J and Zhou, Huaxue and Huang, Yulong and Jin, Kui and Zhou, Fang and Dong, Xiaoli and Zhao, Zhongxian},
  journal={J. Phys. Condens. Matter},
  volume={29},
  number={2},
  pages={025701},
  year={2017},
  doi={http://iopscience.iop.org/0953-8984/29/2/025701}
}

@article{PhysRevLett.73.2364,
  title = {Angular Dependence of the Upper Critical Field of the Heavy Fermion Superconductor ${\mathrm{UPt}}_{3}$},
  author = {Keller, N. and Tholence, J. L. and Huxley, A. and Flouquet, J.},
  journal = {Phys. Rev. Lett.},
  volume = {73},
  issue = {17},
  pages = {2364--2367},
  numpages = {0},
  year = {1994},
  month = {Oct},
  publisher = {American Physical Society},
  doi = {10.1103/PhysRevLett.73.2364},
  url = {https://link.aps.org/doi/10.1103/PhysRevLett.73.2364}
}

@article{PhysRevB.106.134515,
  title = {Two-dimensional multigap superconductivity in bulk $2H\text{\ensuremath{-}}\mathrm{TaSeS}$},
  author = {Patra, C. and Agarwal, T. and Chaudhari, Rajeshwari R. and Singh, R. P.},
  journal = {Phys. Rev. B},
  volume = {106},
  issue = {13},
  pages = {134515},
  numpages = {7},
  year = {2022},
  month = {Oct},
  publisher = {American Physical Society},
  doi = {10.1103/PhysRevB.106.134515},
  url = {https://link.aps.org/doi/10.1103/PhysRevB.106.134515}
}

@article{PhysRevB.83.174506,
  title = {Upper critical field and its anisotropy in LiFeAs},
  author = {Zhang, J. L. and Jiao, L. and Balakirev, F. F. and Wang, X. C. and Jin, C. Q. and Yuan, H. Q.},
  journal = {Phys. Rev. B},
  volume = {83},
  issue = {17},
  pages = {174506},
  numpages = {5},
  year = {2011},
  month = {May},
  publisher = {American Physical Society},
  doi = {10.1103/PhysRevB.83.174506},
  url = {https://link.aps.org/doi/10.1103/PhysRevB.83.174506}
}

@article{kurita2010determination,
  title={Determination of the upper critical field of a single crystal LiFeAs: The magnetic torque study up to 35 Tesla},
  author={Kurita, Nobuyuki and Kitagawa, Kentaro and Matsubayashi, Kazuyuki and Kismarahardja, Ade and Choi, Eun-Sang and S. Brooks, James and Uwatoko, Yoshiya and Uji, Shinya and Terashima, Taichi},
  journal={J. Phys. Soc. Jpn.},
  volume={80},
  number={1},
  pages={013706},
  year={2010},
  publisher={The Physical Society of Japan},
  doi={https://doi.org/10.1143/JPSJ.80.013706}
}

@article{SOTO2007789,
title = {Electric and magnetic characterization of NbSe$_2$ single crystals: Anisotropic superconducting fluctuations above TC},
journal = {Physica C},
volume = {460-462},
pages = {789-790},
year = {2007},
issn = {0921-4534},
doi = {https://doi.org/10.1016/j.physc.2007.04.032},
url = {https://www.sciencedirect.com/science/article/pii/S0921453407005667},
author = {F. Soto and H. Berger and L. Cabo and C. Carballeira and J. Mosqueira and D. Pavuna and P. Toimil and F. Vidal},
}

@article{doi:10.1021/jacs.3c11968,
author = {Cheng, Jingwen and Bai, Jianli and Ruan, Binbin and Liu, Pinyu and Huang, Yu and Dong, Qingxin and Huang, Yifei and Sun, Yingrui and Li, Cundong and Zhang, Libo and Liu, Qiaoyu and Zhu, Wenliang and Ren, Zhian and Chen, Genfu},
title = {Superconductivity in a Layered Cobalt Oxychalcogenide Na$_2$CoSe$_2$O with a Triangular Lattice},
journal = {J. Am. Chem. Soc.},
volume = {146},
number = {9},
pages = {5908-5915},
year = {2024},
doi = {10.1021/jacs.3c11968}
}

@article{KELLER1995568,
title = {Review on the phase diagram and the upper critical field anisotropy of uranium-based heavy fermion superconductors},
journal = {Physica B: Condens. Matter},
volume = {206-207},
pages = {568-573},
year = {1995},
note = {Proceedings of the International Conference on Strongly Correlated Electron Systems},
issn = {0921-4526},
doi = {https://doi.org/10.1016/0921-4526(94)00521-V},
url = {https://www.sciencedirect.com/science/article/pii/092145269400521V},
author = {N. Keller and J.P. Brison and P. Lejay and J.L. Tholence and A. Huxley and L. Schmidt and A. Buzdin and J. Flouquet},
}

@article{Scheidt_2015,
doi = {10.1088/0953-8984/27/15/155701},
url = {https://dx.doi.org/10.1088/0953-8984/27/15/155701},
year = {2015},
month = {mar},
publisher = {IOP Publishing},
volume = {27},
number = {15},
pages = {155701},
author = {Scheidt, E-W and Herzinger, M and Fischer, A and Schmitz, D and Reiners, J and Mayr, F and Loder, F and Baenitz, M and Scherer, W},
title = {On the nature of superconductivity in the anisotropic dichalcogenide NbSe$_2$\{CoCp$_2$\}$_x$},
journal = {J. Phys. : Condens. Matter},
}

@article{PhysRevB.98.035203,
  title = {Enhanced superconductivity upon weakening of charge density wave transport in $2H{\text{-TaS}}_{2}$ in the two-dimensional limit},
  author = {Yang, Yafang and Fang, Shiang and Fatemi, Valla and Ruhman, Jonathan and Navarro-Moratalla, Efr\'en and Watanabe, Kenji and Taniguchi, Takashi and Kaxiras, Efthimios and Jarillo-Herrero, Pablo},
  journal = {Phys. Rev. B},
  volume = {98},
  issue = {3},
  pages = {035203},
  numpages = {9},
  year = {2018},
  month = {Jul},
  publisher = {American Physical Society},
  doi = {10.1103/PhysRevB.98.035203},
  url = {https://link.aps.org/doi/10.1103/PhysRevB.98.035203}
}

@article{abdel2016enhancement,
  title={Enhancement of superconductivity under pressure and the magnetic phase diagram of tantalum disulfide single crystals},
  author={Abdel-Hafiez, M and Zhao, X-M and Kordyuk, Alexander A and Fang, Y-W and Pan, B and He, Z and Duan, C-G and Zhao, J and Chen, X-J},
  journal={Sci. Rep.},
  volume={6},
  number={1},
  pages={31824},
  year={2016},
  publisher={Nature Publishing Group UK London},
  doi={https://doi.org/10.1038/srep31824}
}

@article{PhysRevB.78.224512,
  title = {Transport and anisotropy in single-crystalline ${\text{SrFe}}_{2}{\text{As}}_{2}$ and ${A}_{0.6}{\text{K}}_{0.4}{\text{Fe}}_{2}{\text{As}}_{2}$ ($A=\text{Sr}$, Ba) superconductors},
  author = {Chen, G. F. and Li, Z. and Dong, J. and Li, G. and Hu, W. Z. and Zhang, X. D. and Song, X. H. and Zheng, P. and Wang, N. L. and Luo, J. L.},
  journal = {Phys. Rev. B},
  volume = {78},
  issue = {22},
  pages = {224512},
  numpages = {6},
  year = {2008},
  month = {Dec},
  publisher = {American Physical Society},
  doi = {10.1103/PhysRevB.78.224512},
  url = {https://link.aps.org/doi/10.1103/PhysRevB.78.224512}
}
\end{document}